\documentstyle[aps]{revtex}
\begin{document}
\draft
\title{
Effects of friction on the chiral symmetry restoration 
in high energy heavy-ion collisions
}
\author{Masamichi Ishihara
\thanks{Electric address: m\_isihar@nucl.phys.tohoku.ac.jp}
and 
Fujio Takagi
\thanks{Electric address: takagi@nucl.phys.tohoku.ac.jp}
}
\address{Department of Physics, Tohoku University \\
Aoba-ku, Sendai 980-8578, Japan 
}
\date{\today}
\maketitle
\begin{abstract}
We study the effects of friction on the chiral symmetry restoration  
which may take place temporarily in high energy heavy ion collisions.
The equations of motion with friction are introduced 
to describe the time evolution of the chiral condensates  
within the framework of the linear $\sigma$ model.
Four types of friction are used to study how the result is sensitive
to the choice of the friction.
For the thermalization stage,
the time dependent temperature is parameterized 
so as to simulate the result of the parton-cascade model.
It is parameterized according to the one dimensional scaling hydrodynamics 
for the subsequent cooling stage.
The time development of the condensates and the entropy production due to 
friction are calculated numerically.
The time interval in which the chiral symmetry is restored approximately 
is investigated  in detail for four types of friction.
It is found that; 
(i) the maximum temperature must be high enough (not lower than 230 MeV) 
and the friction must be strong enough 
in order that the chiral symmetry restoration lasts for a long time 
(not shorter than 3fm/c); 
(ii) the ratio of time interval in which chiral symmetry is restored, 
to the time interval in which the temperature is higher than the critical temperature
is typically 0.5 when the friction is strong enough;  
and (iii)
the entropy due to the friction is mainly produced in the early stage of the cooling.
The effect of freezeout is discussed briefly. 
\end{abstract}

\section{Introduction}
\label{sec:intro}
A new phase of matter called quark-gluon-plasma (QGP) is expected to be  
produced at high energy heavy ion collisions. 
The chiral symmetry may be restored in the new phase.
It is usually claimed that chiral symmetry is restored, 
i.e. the chiral condensates vanish
when the temperature $T$ of the system is higher than the critical temperature $T_{c}$.
However, this is true only when the system is always near the equilibrium.
The effective potential may be chirally symmetric 
if $T \ge  T_{c}$.
However, the condition $T \ge T_{c}$ does not necessarily guarantee the vanishing of 
chiral condensates in a time dependent process because of a finite 
relaxation time.
In our previous work \cite{Ishihara1} 
where friction is not taken into account, 
it was found that the maximum temperature must be high enough for the chiral
condensates to vanish temporarily and 
furthermore the chiral condensates cannot stay near the origin of the chiral space
for a long time in general.
The reason is as follows.
For $T \ge T_{c}$,
the time derivative of the chiral condensates becomes maximum 
at the bottom of the single-well potential if there is no friction.
Then, it is difficult for the condensates to stay near the minimum of the potential 
for a long time.
This implies that 
the chiral symmetry does not restore for a long time even 
when $T$ becomes much higher than $T_{c}$.
It may, however, be possible that the condensates stay near the minimum of the potential 
if there is friction of sufficient magnitude\cite{Biro,Rischke,Yabu,masatsu}. 
Therefore, it is important to study the effect of the friction 
on the motion of the condensates in the chiral space. 
Here, we consider that the vanishing or very small chiral condensates are the 
signal of the chiral symmetry restoration.

The chiral symmetry restoration is directly related to the interesting phenomena 
called disoriented chiral condensates (DCC)
\cite{Biro,Rischke,Yabu,masatsu,quench,Asakawa,Gavin,DCCs}.
In the conventional scenarios of DCC formation, 
it is usually assumed that 
the condensates roll down to the minimum of the temperature dependent 
effective potential from the top of the hill of the potential.
In such scenarios, it is implicitly assumed that 
the chiral symmetry restoration takes place for a sufficiently long time.
Then, it is obvious that the investigation of the chiral symmetry restoration 
in a dynamical process is important also for studying DCC formation.

In this paper, we investigate whether the chiral condensates can stay 
near the origin of the chiral space for a long time 
when the equation of motion has a friction term
using the linear $\sigma$ model 
with the massless free particle approximation (MFPA) \cite{Gavin}. 
The conditions of high energy heavy ion collisions are taken into 
account through an appropriate initial condition and a time dependent temperature.

This paper is organized as follows. 
In the next section, 
we derive the equations of motion for the chiral condensates.
Various types of friction are introduced phenomenologically
to study its role played in the chiral symmetry restoration
in heavy ion collision processes.
An expression for the entropy production due to the friction is derived.
In section \ref{sec:numerical},
the equations of motion are solved numerically for an adequate initial condition and 
a time dependent temperature.
The time interval in which the chiral symmetry is restored approximately
is studied in detail. Entropy production is also studied.
Section \ref{sec:summary} is devoted to conclusions and discussions.

\section{The linear $\sigma$ model and the equation of motion with friction}
\label{sec:derive}
\subsection{Equation of motion}
The linear $\sigma$ model is a useful tool to investigate the time evolution of 
the chiral condensates. The Lagrangian is 
\begin{equation}
{\cal L} = \frac{1}{2}\partial_{\mu} \phi \partial^{\mu} \phi  
 - \frac{\lambda}{4}\left(\phi^{2} -v^{2} \right) ^{2} + H \sigma ,
\label{eqn:Lagrangian}
\end{equation}
where $\phi = (\sigma,\vec{\pi})$ and  
$\phi^{2} = \left( \phi_{0},\phi_{1},\phi_{2},\phi_{3} \right) = \sum_{j=0}^{N-1} \phi_{j}^{2}$. 
The equation of motion for the field $\phi_{i}$ is 
\begin{equation}
\Box \phi_{i} + \lambda \left(\phi^{2} -v^{2} \right) \phi_{i} - H \delta_{i,0} = 0 .
\end{equation}
The field $\phi_{i}$ is divided into two parts, 
the condensate $\Phi_{i} = \langle \phi_{i} \rangle$ which is the expectation value of 
the field $\phi_{i}$
and the fluctuation $\tilde{\phi}_{i} = \phi_{i} - \Phi_{i}$. 
Then, we obtain
\begin{eqnarray}
&& \Box \Phi_{i} + \lambda \left(\Phi^{2} -v^{2} \right) \Phi_{i} - H \delta_{i,0} 
+ \Box \tilde{\phi}_{i}  + m_{i}^{2}  \tilde{\phi}_{i} 
+ \lambda \left[  \tilde{\phi}^{2} + 2 \sum_{j \neq i} \Phi_{j} \tilde{\phi}_{j} \right] \Phi_{i} 
\nonumber \\ && \hspace{5cm}
+ \lambda \left[ \tilde{\phi}^{2} + 2 \sum_{j} \Phi_{j} \tilde{\phi}_{j} \right] \tilde{\phi}_{i} = 0 ,
\label{eqn:complex-eq}
\end{eqnarray}
where 
$m_{i}^{2} = 
2 \lambda \Phi_{i}^{2} + \lambda \left( \Phi^{2} - v^{2} \right)$ . 

In eq.(\ref{eqn:complex-eq}), we take  
the normal ordering with respect to the vacuum for which 
the expectation value of $\phi_{i}$ is $\Phi_{i}$.
Assuming that the fluctuation fields $\tilde{\phi_{i}}$ are in the thermal equilibrium,
we obtain the following equation of motion:
\begin{equation} 
\Box \Phi_{i} + \lambda \left(\Phi^{2} -v^{2} \right) \Phi_{i} - H \delta_{i,0}
+ \lambda \left[ \langle : \tilde{\phi}^{2} : \rangle 
+ 2  \langle : \tilde{\phi}_{i}^{2} : \rangle \right] \Phi_{i} = 0.
\label{eqn:neglect-eq}
\end{equation}
Note that the thermal average of the third order term in $\tilde{\phi}$ 
is vanishing under the free particle approximation.
On the other hand, the thermal average of the $:{\tilde{\phi_{i}}}^{2}:$ 
is given by
\begin{equation}
\langle : \tilde{\phi}_{i}^{2} : \rangle = 
\int \frac{d\vec{k}}{(2\pi)^{3} \omega_{k,i}} 
\left[ 
\frac{1}{\exp \left( \omega_{\vec{k},i} / T \right) - 1}
\right] , 
\label{eqn:2nd}
\end{equation}
where 
$\omega_{k,i} = \sqrt{\vec{k}^{2} + m_{i}^{2}}$ . 
It is equal to $T^{2}/12$ for $m_{i} = 0$. 
As a result, the equation of motion of condensates under MFPA
\cite{Gavin} turns out to be 
\begin{equation}
\Box \Phi_{i} + \lambda \left( \Phi^{2} + \frac{T^{2}}{2} - v^{2} \right) \Phi_{i} - H \delta_{i,0} = 0 .
\label{eqn:eq_under_MFPA}
\end{equation}
One can see from eq.(\ref{eqn:eq_under_MFPA}) that 
the potential for the condensates is given by
\begin{equation}
V(\Phi,\Phi_{0},T) = \frac{\lambda}{4} \left( 
\Phi^{2} - v^{2} + \frac{T^{2}}{2}
\right)^{2} -  H \Phi_{0} .
\label{eqn:effective_potential}
\end{equation}
We define the temperature dependent mass squared by the second derivative of $V(\Phi,\Phi_{0},T)$:
\begin{equation} 
\left[ M_{i}(\Phi,\Phi_{i},T) \right] ^{2} 
= \frac{d^{2} V}{d \Phi_{i}^{2}} 
= \lambda \left( \Phi^{2} - v^{2} + \frac{T^{2}}{2} + 2 \Phi_{i}^{2} \right).
\label{eqn:temperature_dep_mass}
\end{equation} 

\subsection{Choice of friction}
\label{sec:subsec:fric}
As already mentioned in Sec.\ref{sec:intro}, 
the aim of this paper is to investigate the friction effects 
on the chiral symmetry restoration in high energy heavy ion collisions.
The system which consists of the chiral condensates (soft modes) will 
be far from a (local) thermal equilibrium in such a time dependent process 
even when the environment which consists of the fluctuation fields (hard modes) is approximately 
in the (local) thermal equilibrium. 
In such a case, a friction term appears in general as a result of 
the dissipation-fluctuation theorem \cite{fluctu}.
Thus we use a phenomenological equation of motion with a friction term:
\begin{equation}
\Box \Phi_{i} + \lambda \left( \Phi^{2} + \frac{T^{2}}{2} - v^{2} \right) \Phi_{i} - H \delta_{i,0}  
+ \sum_{j} \eta_{ij} \frac{d\Phi_{j}}{dt} = 0.
\label{eqn:biro_type}
\end{equation}
\noindent
Here, $\eta_{ij}$'s are the friction coefficients.
In general, the friction coefficients depend on both $T$ and $\Phi_{i}$. 
In this paper, we assume that 
$\eta_{ij} = \eta \delta_{ij}$. 
For the functional form of $\eta$ which may provide either 
$T$ and/or $\Phi_{i}$ dependence, 
we use the following form which was obtained for the chirally symmetric vacuum 
\cite{Biro,Rischke} (See also, \cite{Joen}):
\begin{equation}
\eta = \frac{9}{16 \pi^{3}} \frac{\lambda^{2} T^{2}}{m} f_{Sp} \left( 1 - e^{-m/T} \right)
    \equiv F(m,T) ,
\label{eqn:cof_expr}
\end{equation}
where $f_{Sp}(x) = - \int_{1}^{x} dt \frac{\ln t}{t-1}$ is the Spence function and 
$m$ is the temperature dependent mass.  
In this paper, 
we consider the following four cases as the choice of $\eta$:

{
\setcounter{enumi}{\value{equation}}
\addtocounter{enumi}{1}
\setcounter{equation}{0}
\renewcommand{\theequation}{\theenumi\alph{equation}}
\noindent 
case A :
\begin{equation}
\eta = {\rm constant.}
\label{eqn:firc:const}
\end{equation}
i.e. $\eta$ is taken for simplicity as a constant 
which is independent of both $T$ and $\Phi_{i}$ 
\cite{Biro,masatsu}

\noindent
case B :
\begin{equation}
\eta = F(m_{\pi},T) ,
\label{eqn:firc:fixes}
\end{equation}
where $m_{\pi}$ is the pion mass at zero temperature.
This choice may serve as a case in which 
$\eta$ has the simplest nontrivial $T$-dependence.
This case may be 
realized in the rapid cooling case.

\noindent
case C :
\begin{equation}
\eta = F \left( M_{3}(f_{\pi},0,T), T \right) .
\label{eqn:firc:effective}
\end{equation}
This is another case where $\eta$ has a simple $T$-dependence. 
In this case, we ignore the change of $\Phi$ in the $\Phi$- and $T$-
dependent mass defined in eq.(\ref{eqn:temperature_dep_mass}), 
i.e., 
we take $\Phi_{0} = f_{\pi}$ and $\Phi_{i} = 0$ for $i=1,2,3$ 
in order that $M_{3}(f_{\pi},0,T)$ coincides with the pion mass at $T = 0$. 


\noindent
case D : 
\begin{equation}
\eta = F \left( M_{3}(\Phi_{v},0,T), T \right),
\label{eqn:firc:condensation}
\end{equation}
where $\Phi_{v}$ is the condensation value of the temperature dependent vacuum 
which is determined by the extremum condition
\begin{equation}
\left. \frac{\partial V}{\partial \Phi_{0}} \right|_{\Phi_{0}=\Phi_{v},\vec{\Phi}=0} =0 .
\end{equation}
In this case, the mass $M_{3}$ depends on both $T$ and $\Phi_{v}$ while the latter is a 
function of $T$. 
Note that only case A has been considered by previous authors in numerical analyses \cite{Biro}.
\setcounter{equation}{\value{enumi}}
}

The magnitude of the friction may be estimated
from eq.(\ref{eqn:cof_expr}) for the symmetric phase.
Bir\'o and Greiner suggested that $\eta = 2.2 {\rm fm}^{-1}$ 
at $T \sim T_{c}$ \cite{Biro}.
Rischke calculated $\eta$  for the broken phase  \cite{Rischke}.
It was also estimated by Yabu et al. \cite{Yabu} using the Caldeira-Legegett method. 
Their result is $\eta = 0.7 m_{\pi} $ in the uniform case.
For the case A, 
we use $\eta = 2 {\rm fm}^{-1}$ as a typical value and consider also some other values for comparison. 
Considering the theoretical ambiguity in the magnitude of $\eta$, 
we introduce the overall  multiplicative constant $C_{\eta}$ for cases B,C and D.
The friction coefficient $\eta$ in these three cases are replaced with $C_{\eta} \eta$ 
with various $C_{\eta}$ in the following numerical calculations.

\subsection{Time dependence of temperature}

We consider the time development of chiral condensates in the entire stage of 
heavy ion collisions at high energies. 
We concentrate on the central rapidity region where the temperature will become 
highest. 
A process which takes place in this region will be approximately invariant 
under the Lorentz boost along the collision axis. 
Then a convenient variable is the proper time $\tau$. 
The condensates which are formed in the central region will evolve under the influence of the 
environment which consists of many number of partons (quarks, antiquarks and gluons) or hadrons. 
The effect may be represented by the temperature $T$ which depends on  the proper time $\tau$ 
provided the environment is in the local thermal equilibrium
\cite{Ishihara1}.
We use the temperature as a convenient parameter even for the very early stage of the collision
process where the environment may be far from the thermal equilibrium. 
This is because we do not know any other reliable method to describe the effect of the environment 
in the pre-equilibrium stage.

It is conceivable that hard collisions among the incident partons play a crucial role to 
thermalize the system in the beginning of the entire process. 
We then borrow the result of the parton cascade model \cite{Geiger}
in order to parameterize the $\tau$-dependence of $T$ for the early themalization stage
including the very early stage of the collision. 
It is possible to estimate the $\tau$ - dependent temperature $T(\tau)$ up to 
$\tau = \tau_{m}$ when the temperature reaches the maximum $T_{m}$ using the energy densities 
calculated in $\cite{Geiger}$ with the equation of state for the free quark-gluon gas. 
For example, one has $T_{m} \sim 620$ MeV and $\sim 880$ MeV for 
Au-Au collisions at RHIC and LHC energies, respectively.
The proper time $\tau_{m}$ is typically some 1 fm. 
After this time, the system will expand mainly along the collision axis. 
The one-dimensional scaling hydrodynamics 
\cite{Bjorken}
will be a good approximation for the cooling stage.
Then the temperature becomes a function of $\tau$ only, 
$T=T(\tau) \propto \tau^{-1/3}$  for $\tau \geq \tau_{m}$. 
To summarize, the temperature is parameterized as follows as a function of the 
scaled proper time $x = \tau/\tau_{m}$ for the entire stage of the collision 
\cite{Ishihara1} :
\begin{equation}
T(x) = T_{i} \left( \frac{T_{m}}{T_{i}} \right)^{x} \theta(1-x) 
     + T_{m} x^{-1/3} \theta(x-1) .
\label{eqn:temperature}
\end{equation}

The temperature parameterized in eq.(\ref{eqn:temperature}) stays finite for a long time.
However, the interactions between the environment and the condensates will vanish 
at a certain temperature called the freezeout temperature $T_{f}$. 
Therefore, it is reasonable to consider that $T$ suddenly becomes zero when it reaches $T_{f}$.
Then, the time dependence of the temperature may be parameterized as
\begin{equation}
T(x) = T_{i} \left( \frac{T_{m}}{T_{i}} \right)^{x} \theta(1-x) 
     + T_{m} x^{-1/3} \theta(x-1) \theta(x_{f} -x) ,
\label{eqn:temperature_freezeout}
\end{equation}
where $x_{f} = \left( T_{m}/T_{f} \right)^{3} $ and  
$\tau_{f} = x_{f} \tau_{m} $ being the freezeout time.
To summarize, the temperature decreases from $T_{m}$ down to $T_{f}$ and then 
suddenly becomes zero. 
It is supposed that the friction vanishes after $x_{f}$ because there will be 
no energy dissipation from the condensates  into the thermal environment after  
the freezeout. 
In this situation, the initial condition of the quench scenario 
for the formation of DCC 
\cite{quench,Asakawa}
may be realized at $x = x_{f}$. 
On the other hand, 
a whole process may exhibit a characteristic feature of the annealing scenario
\cite{Asakawa,Gavin}
if $T_{f}$ is much lower than $T_{c}$ and if the one dimensional expansion lasts 
for a long time. 
Equation of motion, eq.(\ref{eqn:biro_type}), is now rewritten as 
\begin{equation}
\left[
\frac{\partial^{2}}{\partial \tau^{2}} + \frac{1}{\tau} \frac{\partial}{\partial \tau} 
+ \lambda \left(\Phi^{2} + \frac{T^{2}(x) }{2} - v^{2} \right) 
\right]
\Phi_{i} 
- H \delta_{i,0}  + \eta \frac{\partial \Phi_{i}}{\partial \tau} = 0 ,
\label{eqn:1dim_normal}
\end{equation}
where we assume that the system is homogeneous along the transverse direction. 
In the next section, eq.(\ref{eqn:1dim_normal}) will be solved numerically 
with $T(x)$ given by 
eq.(\ref{eqn:temperature}) or eq.(\ref{eqn:temperature_freezeout}).

\subsection{Entropy production}
Consider the following equation of motion of a particle with mass $m$ in order to 
find the expression for the  entropy production:
\begin{equation}
m \frac{d^{2} X}{dt^{2}} = F(X) + G (X) , 
\end{equation}
where $F(X)$ represents a conservative force and $G(X)$  a non-conservative one.
The work $W$ which the particle does by $G(X)$ is 
\begin{equation}
W = - \int_{\rm c} dX G(X),
\end{equation}
where ${\rm c}$ is the path of the particle . 
If the work $W$ is fully converted into heat, 
the produced entropy is given by
\begin{equation}
S = \int \frac{dQ}{T} = - \int_{c} dX \frac{G(X)}{T} 
= - \int_{\rm c} dt \frac{dX}{dt} \frac{G(X)}{T}  .
\end{equation}
In the present case, 
$X$ corresponds to $\Phi_{i}$ and 
$G(X)$ to $- \sum_{j} \eta_{ij}(\Phi) d\Phi_{j}/dt$ in eq.(\ref{eqn:biro_type}).
Then, The increase in the entropy density may be given by
\begin{equation}
ds_{i} = \sum_{j} d\Phi_{i} \frac{\eta_{ij}}{T(x)} \left( \frac{d\Phi_{j}}{d\tau} \right)
   = d\tau \sum_{j} \frac{\eta_{ij}}{T(x)} 
   \left( \frac{d \Phi_{i}}{d\tau} \right)\left( \frac{d \Phi_{j}}{d\tau} \right), 
\end{equation}
where $t$  has been replaced with $\tau$. 
The above expression is in agreement with eq.(4.6) in ref.\cite{Hosoya}
when there is only one kind of field.
The entropy density production per unit proper time is given by 
\begin{equation}
\frac{ds_{i}}{d\tau} =  \sum_{j} \frac{\eta_{ij}}{T(x)} 
\left(\frac{d {\Phi}_{i}}{d \tau} \right) \left(\frac{d {\Phi}_{j}}{d \tau} \right) . 
\label{eqn:dsdtau}
\end{equation}

\section{Numerical Results}
\label{sec:numerical}

\subsection{Choice of parameters}
We take the parameters of the linear $\sigma$ model as follows 
in the following numerical calculations:
$\lambda = 20$, $v = 87.4$ 
and $H^{1/3}$ = 119 MeV.
This set of the parameters generates the pion mass 135 MeV, 
the sigma mass 600 MeV and the pion decay constant $f_{\pi}$ = 92.5 MeV.
The initial condition of the equation of motion, eq.(\ref{eqn:1dim_normal}),  is taken as 
{
\setcounter{enumi}{\value{equation}}
\addtocounter{enumi}{1}
\setcounter{equation}{0}
\renewcommand{\theequation}{\theenumi\alph{equation}}
\begin{eqnarray}
\left(\Phi_{0}, \vec{\Phi} \right) &=& \left( f_{\pi},\vec{0} \right), 
\\
\left. \left( \frac{d\Phi_{0}}{d\tau} , \frac{d \Phi_{1} }{d\tau} 
, \frac{d \Phi_{2} }{d\tau} , \frac{d \Phi_{3} }{d\tau} \right) \right|_{\tau = 0}
&=&
\left( - \frac{\pi T_{\rm i}^{2}}{\sqrt{30}},
  \frac{\pi T_{\rm i}^{2}}{\sqrt{30}} , 
  \frac{\pi T_{\rm i}^{2}}{\sqrt{30}} , 
  \frac{\pi T_{\rm i}^{2}}{\sqrt{30}} \right), 
\end{eqnarray}
\setcounter{equation}{\value{enumi}}
}
where $T_{\rm i}$ is the initial temperature.

Four types of friction given in eqs. 
(\ref{eqn:firc:const}) $\sim$ (\ref{eqn:firc:condensation})
are used in the following calculations. 
The $T$-dependence of the friction in cases B,C and D
is displayed in Figs.\ref{fig:Frictions1} and \ref{fig:Frictions2}.
The $T$-dependence in the case D comes from the $T$-dependence of the pion mass. 
The pion mass is almost constant below $T_{c}$ and increases linearly above $T_{c}$.   
This behavior of the pion mass below $T_{c}$ is similar to that in case B.
The behavior above $T_{c}$ is asymptotically identical to that in case C.
Moreover, the magnitude of the friction is determined mainly by the factor $T^{2}/m(T)$. 
As a result, the $T$-dependence of the friction in case D is quite similar
to that in  case B at low temperatures and 
it is similar to that in  case C at high temperatures.
The $T$-dependence of the friction in  case D has a peak near the $T_{c}$
because of $T$-dependence of the pion mass.
Since the $T$-dependence of the friction comes from 
the $T$-dependence of the pion mass,
that in case B is proportional to $T^{2}$, 
while that in case C and D is proportional to $T$ at high temperatures.

\subsection{Time development of chiral condensates}
We have solved eq (\ref{eqn:1dim_normal}) numerically for various types of friction. 
The result for case A with the maximum temperature $T_{\rm m} = 200  {\rm MeV}$, 
the initial temperature $T_{\rm i} = 1 {\rm MeV}$ , 
the time $\tau_{m} = 1 {\rm fm}$ and various $\eta$ 
is shown in Fig.\ref{fig:FricT200Tf1}.
It is found that 
a) the motion of the condensates for $x < 1$ is insensitive to the value of $\eta$
, and 
b) the condensates cannot stay for a sufficiently long time in the center of the chiral space.
As already found in \cite{Ishihara1}, 
a characteristic damped oscillation appears in the no friction case ($\eta = 0$).
Contrary to this, 
the oscillation diminishes rapidly when $\eta$ = 1 and 2 ${\rm fm}^{-1}$. 
The condensates ($\Phi_{0}$) in these two cases 
pass near the center of the oscillation in the $\eta = 0$ case. 

The time development of condensates in cases B,C and D with  
$T_{m} = 250$ MeV, $T_{i} = $1 MeV and $\tau_{m} = $1fm is shown in Fig.\ref{fig:masstype}.
There are characteristic oscillations in cases C and D, while there is not in case B. 
The oscillation in case D is weaker than that in case C. 
These different behaviors reflect the different $T$-dependence of the friction 
in the three cases.

The $T_{m}$ dependence of the condensate $\Phi_{0}$ 
in case A is shown in Fig.\ref{fig:Tmdep_Biro}, 
where the parameters are taken as $T_{i} = 1$ MeV, $\tau_{m} = 1$ fm 
and $\eta = 2 {\rm fm}^{-1}$.
Higher the maximum temperature is, faster it moves to the origin. 
However, it is difficult to be stopped by friction as the condensate moves rapidly. 
The different behavior for $x \ge 10$ is due to the difference of $T$.
Since the behavior of $T$ is determined by the scaling property, 
$T$ depends on only $T_{m}$ if $\tau_{m}$ is fixed. 
Higher $T$ is, smaller the minimum of the potential (eq.(\ref{eqn:effective_potential})) is.
As a result, the condensate is small for high $T_{m}$.

The sensitivity of the time development to the initial temperatures $T_{i}$
is shown in Fig.\ref{fig:InitTdep}.
It is found that the  $T_{i}$ dependence is weak. 

The proper time $\tau_{m}$ is 1 fm or less 
according to the result of the parton cascade model.
The condensates evolves in a different way for different $\tau_{m}$ even if 
other parameters, $T_{m}, T_{i}$ and $\eta$ are fixed. 
Since $\tau_{m}$ determines the behavior of the temperature in the entire stage, 
we show typical $\tau_{m}$ dependence in Fig.\ref{fig:TFdep_Sc_Biro}
for $T_{\rm m} = 250 {\rm MeV}$, $T_{\rm i} = 1 {\rm MeV}$ and $\eta = 2 {\rm fm}^{-1}$.
For  $x < 10$, the time development in $x$ is faster for larger $\tau_{m}$.
On the other hand, the behavior for $x \ge 14$ is almost independent of $\tau_{m}$. 
That is, the behavior at large $x$ scales in $x$. 
The motion at large $x$ is determined solely by $T$ for given $T_{m}, T_{i}$ and $\eta$. 

The effect of freezeout on the time development of condensates in case D is 
demonstrated in Fig.\ref{fig:entrory_massT}.
The curves $(\alpha)$ and $(\beta)$ represent the results obtained 
with the time dependent temperature (\ref{eqn:temperature}) and (\ref{eqn:temperature_freezeout}),
respectively. 
Of course, the effect of freezeout is  seen only for $x > x_{f}$. 
For comparison, $T(x)$ given by eq.(\ref{eqn:temperature_freezeout}) 
is also shown in Fig.\ref{fig:entrory_massT}. 
The center of the oscillation of sigma condensate in the curve ($\beta$) 
is almost $f_{\pi}$.
The amplitude of the oscillation is determined by the magnitude of 
the condensate at the freezeout time. 

\subsection{Chiral symmetry restoration time}
One may claim that the chiral symmetry is restored approximately if every 
$\mid \Phi_{i} \mid$ becomes much smaller than $v$. 
For definiteness, we use the following condition to define the approximate 
restoration of chiral symmetry:
\begin{equation}
\sqrt{\Phi^{2}} / v  \le 0.2 .
\label{eqn:cond}
\end{equation}
From now on, we say that chiral symmetry is restored 
if the condition (\ref{eqn:cond}) is satisfied.

We take $T_{i}$ = 1 MeV and $\tau_{m}$ = 1 fm in the following calculations.  
Time interval $\tau_{r}$ in which chiral symmetry is approximately restored 
is calculated for the four cases with various $T_{m}$.
The result for case A is shown in Fig.\ref{fig:New_biro_time-tf1}.
It is found that the condition $T_{m} > 230$ MeV, $\eta > 1 {\rm fm }^{-1}$ 
must be fulfilled in order that $\tau_{r} \ge 3$ fm.
At fixed $\eta$, higher $T_{m}$ is, longer $\tau_{r}$ becomes. 
This is because 
the condensate follows the minimum of the potential realizing the chirally symmetric vacuum 
and high temperature era lasts for a long time when $T_{m}$ is high.  
On the other hand, for fixed $T_{m}$, 
$\tau_{r}$ increases as a function of $\eta$ for small $\eta$,
reaches a maximum at some $\eta$ and then decreases.
This behavior is understood as follows. 
If $\eta$ is large, the condensate moves slowly and almost stops near the origin 
resulting in a long $\tau_{r}$. 
However, if $\eta$ is too large, $T$ decreases considerably 
before the condensate reaches the origin and hence, $\tau_{r}$ becomes short.
The same tendency is found in other cases.

The result for the case B is shown in Fig.\ref{fig:m135_2} .
Now $\tau_{r}$ is plotted as a function of $C_{\eta}$ 
introduced in subsection \ref{sec:subsec:fric}.
The condition for having $\tau_{r} \ge 3 {\rm fm}^{-1}$  is $T_{m} \ge 250$ MeV 
and $C_{\eta}  \stackrel{>}{\sim} 0.1$.
In this case, large $\tau_{r}$ can be realized even when $C_{\eta}$ is very small 
because $\eta$ becomes very large. (See fig.\ref{fig:Frictions2})
The result for case C is shown in Fig.\ref{fig:effective2}.
The appearance of large $\tau_{r}$ for $T_{m} = 140$ MeV is 
due to an accidental fluctuation.
Maximum temperature must be higher than 260 MeV and  
$C_{\eta}$ must be larger than 2 in order that $\tau_{r} \ge 3$ fm. 
Fig.\ref{fig:VEV2} shows the time interval in case D. 
The condition for having $\tau_{r} \ge 3$ fm  is $T_{m} \ge 220$ MeV and 
$C_{\eta} \stackrel{>}{\sim} 1.5 \sim 2.0$. 
To summarize the results for the four cases, we found the following general tendency
common to every case: 
(a) the $T_{m}$ dependence of $\tau_{r}$ is approximately universal as long as 
the friction is sufficiently but not too strong; 
(b) $T_{m}$ must be larger than 220 $\sim$ 260 MeV in order to have $\tau_{r} \ge 3$ fm;
(c) approximate chiral symmetry restoration does not take place 
    if the friction is very small. 

We define the time interval $\tau_{c}$ in which the temperature of the system is higher 
than the critical temperature $T_{c}$ in order to compare it with $\tau_{r}$. 
It is given by 
\begin{equation}
\tau_{c} \sim \tau_{m} \left( \left[ \frac{T_{m}}{T_{c}} \right]^{3} - 1 \right) .
\end{equation}
This ratio in  case A is shown in Fig.\ref{fig:scale_const}. 
It is found that the typical value of this ratio is about 0.5. 
Each of the lines corresponds to various $T_{m}$ ranging from 230 MeV to 390 MeV.
The ratio is approximately independent of $\eta$ for large $\eta$. 
We have confirmed that these features are common to all the four cases.

\subsection{Entropy production}
Finally, the entropy production due to the friction is calculated. 
The result for case D is already shown in fig.\ref{fig:entrory_massT}.
For comparison, we have calculated it for case A with 
$\eta = 2 {\rm fm}^{-1}$, $T_{\rm m} =$ 250 MeV, 
$T_{\rm i} =$ 1 MeV and $\tau_{m} =$ 1 fm. 
The result is shown in Fig.\ref{fig:ds0dx}.
It is found that entropy is produced mainly in the early cooling stage. 
Entropy production lasts for a relatively short time (a few fm) 
and hence it is not affected by the freezeout. 
The entropy production is mainly due to the motion of $\Phi_{0}$ condensate 

\section{Conclusions and Discussions}
\label{sec:summary}
The time development of chiral condensates in high energy heavy ion collisions 
are calculated with the adequate initial conditions and frictions. 
The linear $\sigma$ model with four types of friction are used.
The main results are: 
a) the characteristic damped oscillation of the condensates 
   in no friction case is smeared by the friction;
b) sufficiently high maximum temperature ($T_{m} \ge 220$MeV) and 
   the friction of appropriate magnitude 
   are needed for the vacuum to be chiral symmetric 
   for a sufficiently long time ($\tau_r \ge 3$fm);
c) entropy is produced mainly in the early stage of the cooling.

For each friction type, we found the following results. 
In case A, the constant friction which has been  used by some authors generates 
the largest $\tau_{r}$ among four cases.
However, it is not clear whether
the $T$ independence of $\eta$ is a good approximation or not. 
In case B, $\tau_{r}$ is long enough even for small $C_{\eta}$. 
However, the magnitude of the friction is unrealistically large 
at high temperatures. 
In case C, $\tau_{r}$ is small for the realistic range of the friction
($ 0.5 \le C_{\eta} \le 3 $) .
In case D, large $\tau_{r}$ is obtained for the possible range of the friction 
($ 0.5 \le C_{\eta} \le 3 $).
The peak of $\eta$ near $T_{c}$ (see Fig.\ref{fig:Frictions1}) is responsible for large $\tau_{r}$ 
in this case.

As stated in the introduction, 
the conventional scenarios of the formation of QGP and DCC in which 
the chiral symmetry is supposed to be restored 
for a sufficiently long time are not always guaranteed. 
Our results suggests that
it is important to estimate precisely  the maximum temperature of the system and 
the friction for the soft mode.

In general scenarios of QGP evolution, 
it is well-known that 
the entropy of the system is mostly  produced in the thermalization stage ($x \le 1 $).
However, we found that  the entropy due to the friction felt by the condensates 
is produced in the early stage of the cooling. 

As the initial condition of conventional DCC formation scenarios,
it is often assumed that the roll-down of chiral condensates starts from the top of 
the potential (chirally symmetric vacuum) 
and the initial temperature is close to $T_{c}$. 
However, this initial condition may not be always satisfied 
as the chiral symmetry restoration does not necessarily occur even when 
the temperature of the system exceeds $T_{c}$.

In our calculation, 
the condensation dependence of the friction is included in case D. 
However, it reflects essentially only the $T$-dependence of the vacuum.
The effect of the time dependent condensates is not taken into account yet.
Such a dependence \cite{Rischke} should be included to refine the analysis.
   On the other hand, random forces are not considered also in this paper 
because we are interested in the average behavior of the chiral condensate. 
 They generate fluctuations in the motion of the condensate and hence in $\tau_{r}$.
The fluctuation-dissipation theorem implies that
the random force is strong when the friction is strong.
However, the strong random force does not always affect considerably 
the motion of the condensate because its effect is reduced by the strong friction.
Even when the random force works, $\tau_{r}$ will become shorter on the average
because condensate will occasionally be kicked out from the chirally symmetric 
region  by the strong random force. 
This means that higher $T_{m}$ is required to realize the same $\tau_{r}$ on 
the average. There is another aspect in this situation.
The condensate in the present case moves mainly along the $\sigma$ axis. 
It is almost at rest along the $\vec{\pi}$ axis.
Though the random force will decelerate or accelerate the condensate along the sigma direction, 
it will only accelerate the condensate along the $\vec{\pi}$ direction.
As a result, the chiral symmetry will be more easily broken when the random force is included.

Finally, we would like to make a comment on the massless free particle approximation.
It is not reliable at low temperature. 
The effects of mass and the finite volume of the colliding system 
have to  be taken into account to improve the theory. 
However,
we believe that our results are correct at least  qualitatively.
The problems listed above and their effects on DCC formation will be 
discussed in the coming paper.


{
\setcounter{enumi}{\value{figure}}
\addtocounter{enumi}{1}
\setcounter{equation}{0}
\renewcommand{\thefigure}{\theenumi\alph{figure}}
\begin{figure}
\caption{The temperature dependence of the  
magnitude of friction in cases B,C and D.}
\label{fig:Frictions1}
\end{figure}


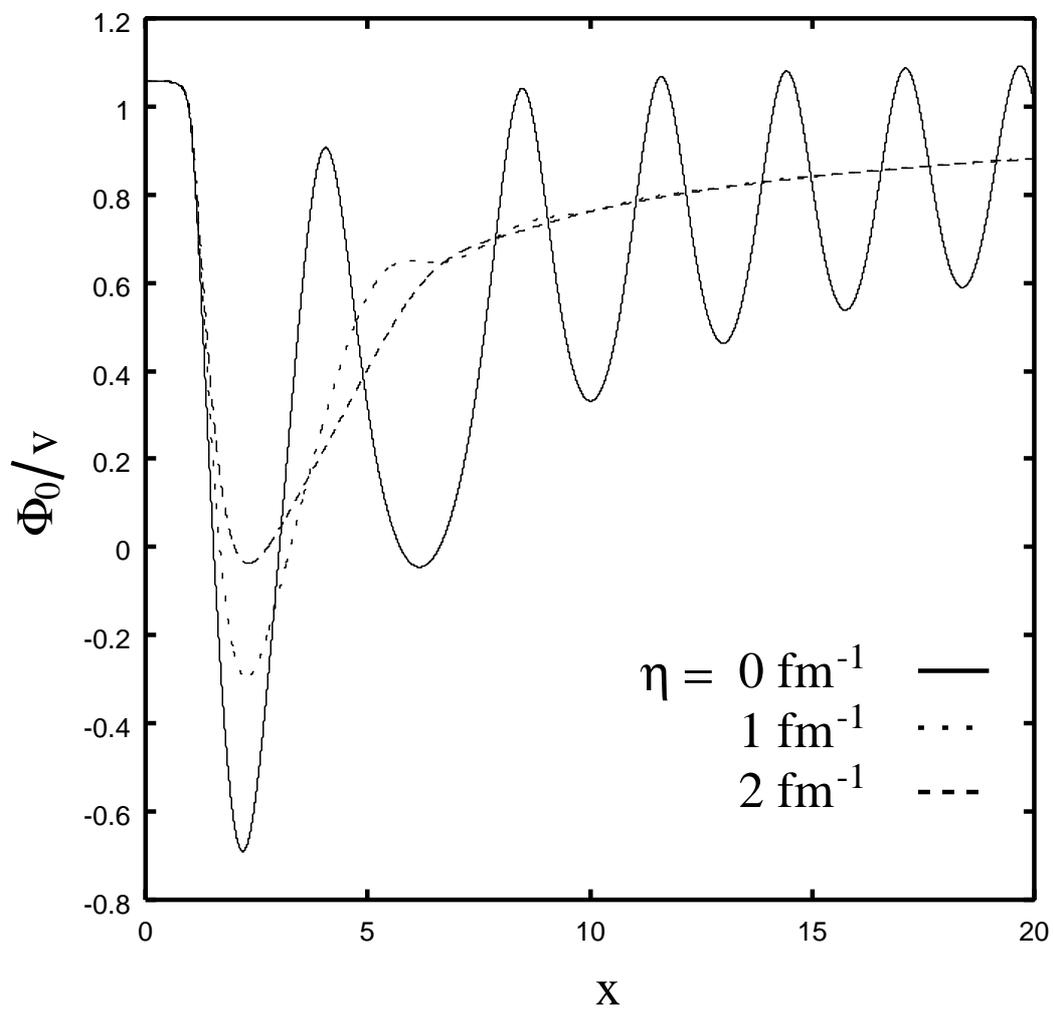
\begin{figure}
\caption{The temperature dependence of the  
magnitude of friction in  cases B,C and D.
Note that the scale of $T$ is different from that in Fig.\ref{fig:Frictions1}.
}
\label{fig:Frictions2}
\end{figure}
\setcounter{figure}{\value{enumi}}
}

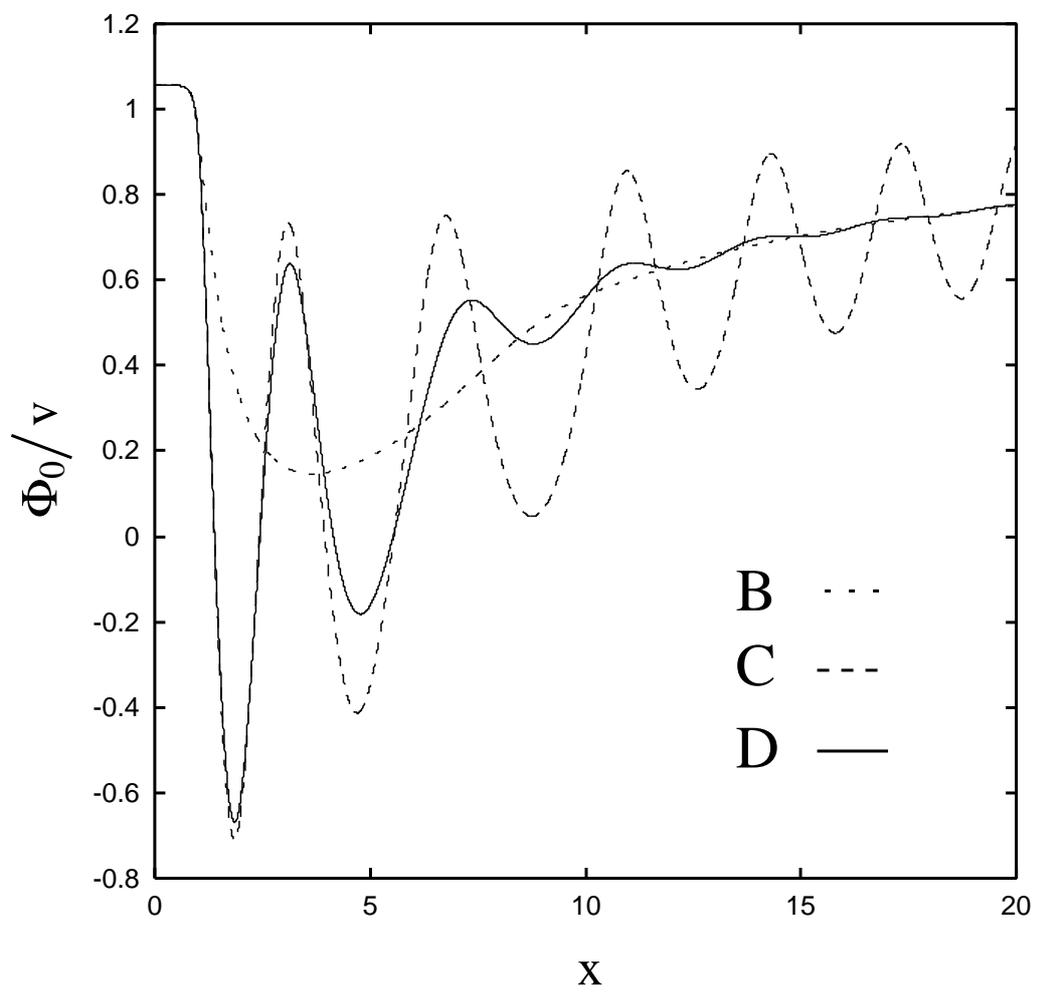
\begin{figure}
\caption{Time development of chiral condensates for various magnitudes of friction
with $T_{m}$ = 200 MeV, $T_{i}$ = 1MeV and $\tau_{m}$ = 1fm in  case A}
\label{fig:FricT200Tf1}
\end{figure}

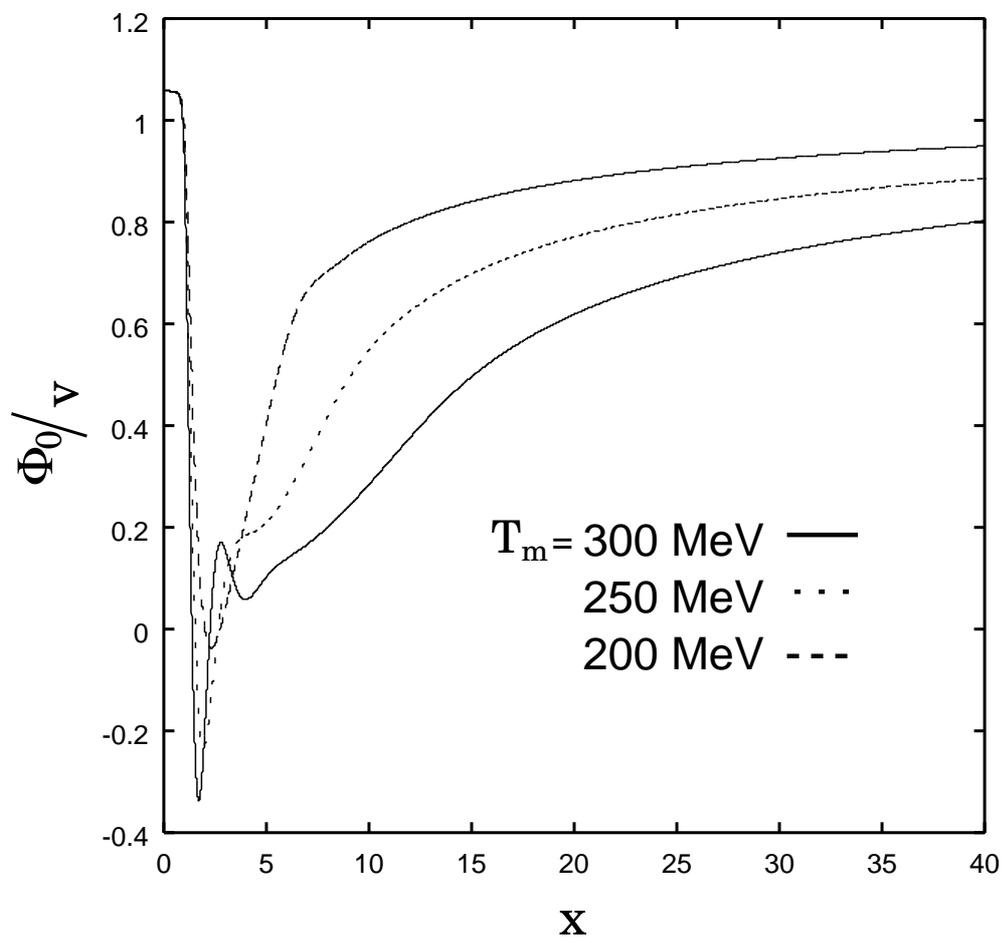
\begin{figure}
\caption{Time development of chiral condensates 
for  $T_{m}$ = 250 MeV, $T_{i}$ = 1MeV and $\tau_{m}$ = 1fm in cases B,C and D.}
\label{fig:masstype}
\end{figure}

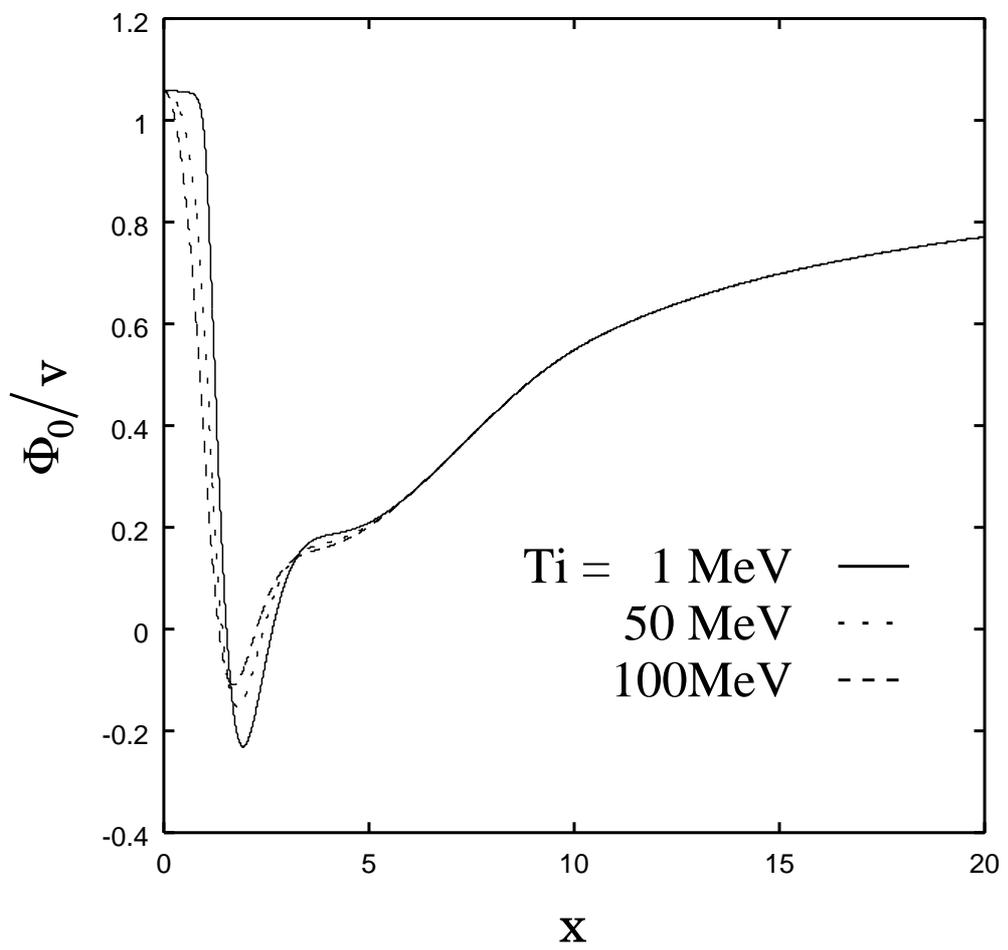
\begin{figure}
\caption{Time development of chiral condensates for various $T_{m}$
with $T_{i}$ = 1MeV, $\tau_{m}$ = 1fm and $\eta = 2{\rm fm}^{-1}$ in  case A.}
\label{fig:Tmdep_Biro}
\end{figure}

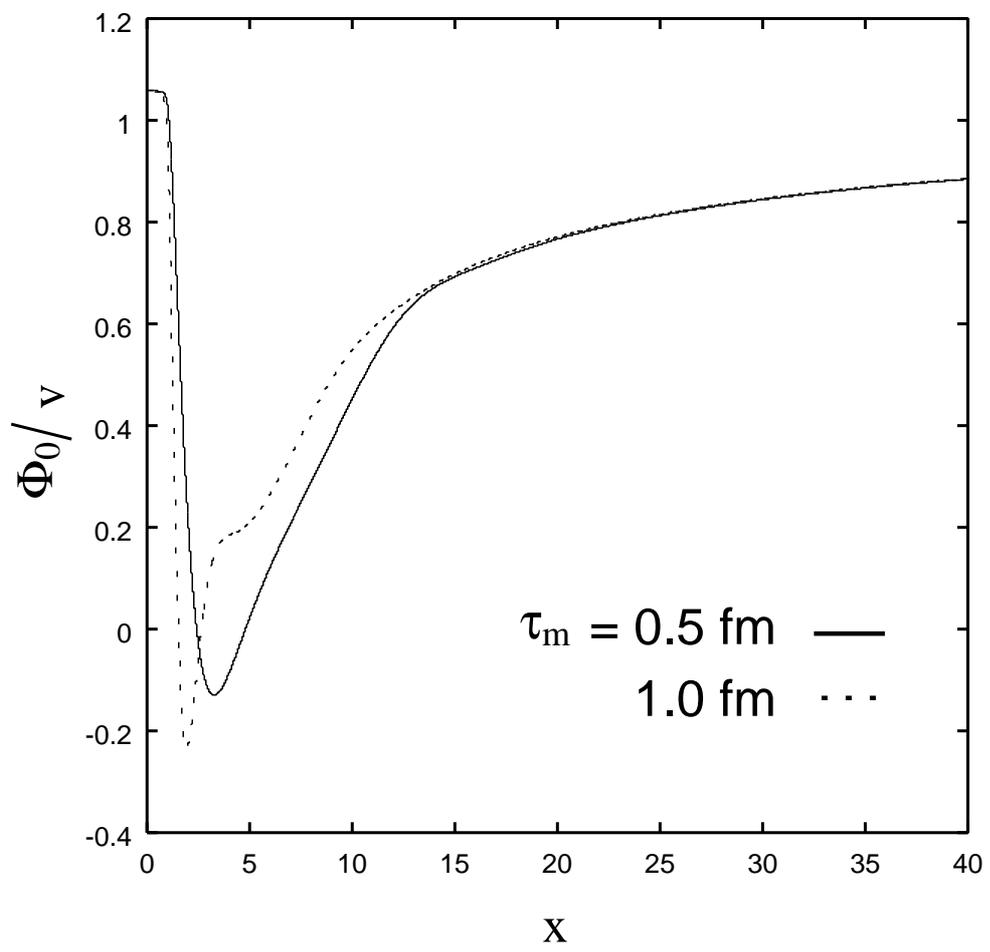
\begin{figure}
\caption{Time development of chiral condensates for various $T_{i}$
with $T_{m}$ = 250 MeV, $\tau_{m}$ = 1fm and $\eta = 2{\rm fm}^{-1}$ in  case A.}
\label{fig:InitTdep}
\end{figure}

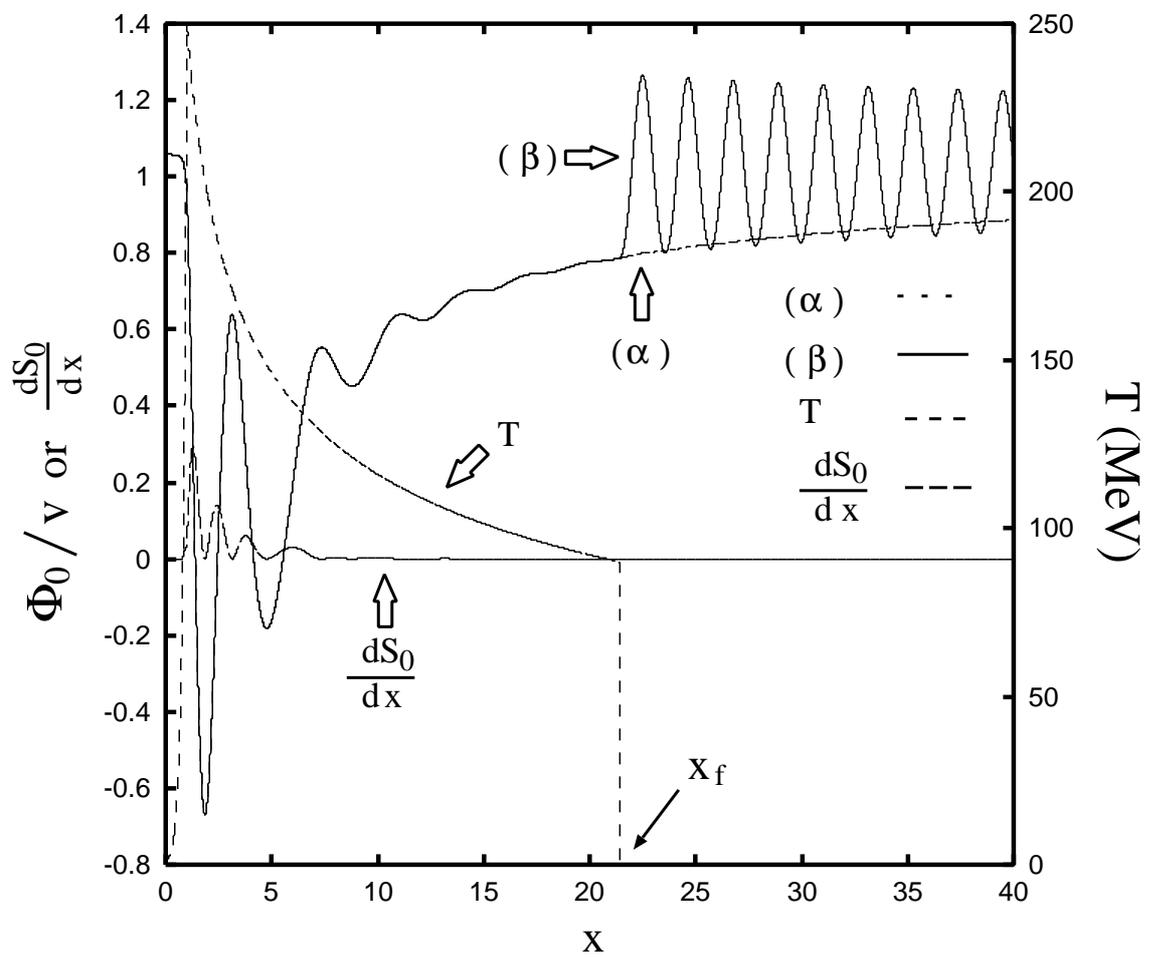
\begin{figure}
\caption{Time development of chiral condensates for various $\tau_{m}$
with $T_{m}$ = 250 MeV, $T_{i}$ = 1MeV and $\eta = 2{\rm fm}^{-1}$ in case A.
}
\label{fig:TFdep_Sc_Biro}
\end{figure}

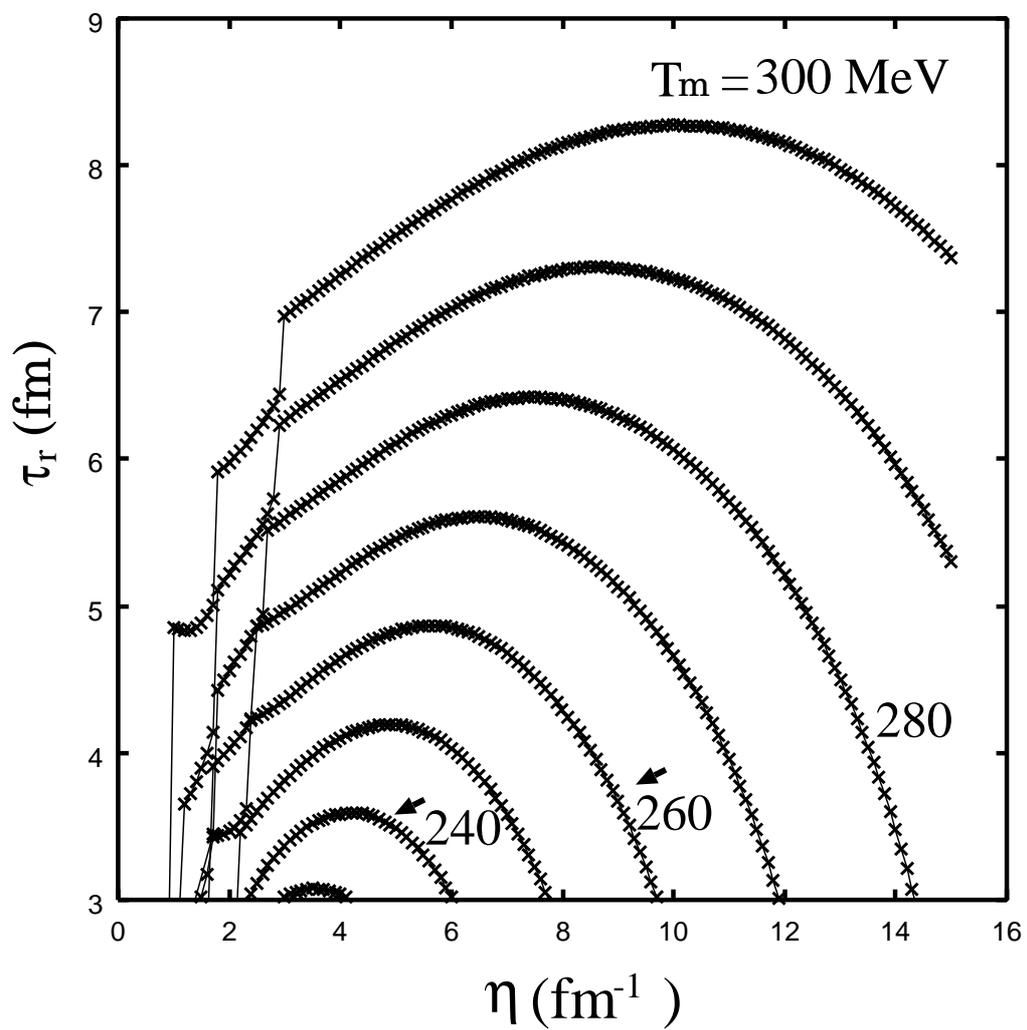
\begin{figure}
\caption{
Time development of sigma condensate and 
entropy production per unit scaled time $(\tau/\tau_{m})$ 
without $(\alpha)$ and with $(\beta)$ the freezeout
in  case D.
The parameters are 
$T_{m}$ = 250 MeV , $T_{i}$ = 1 MeV, $\tau_{m}$ = 1fm and $T_{f}$ = 90MeV.
}
\label{fig:entrory_massT}
\end{figure}

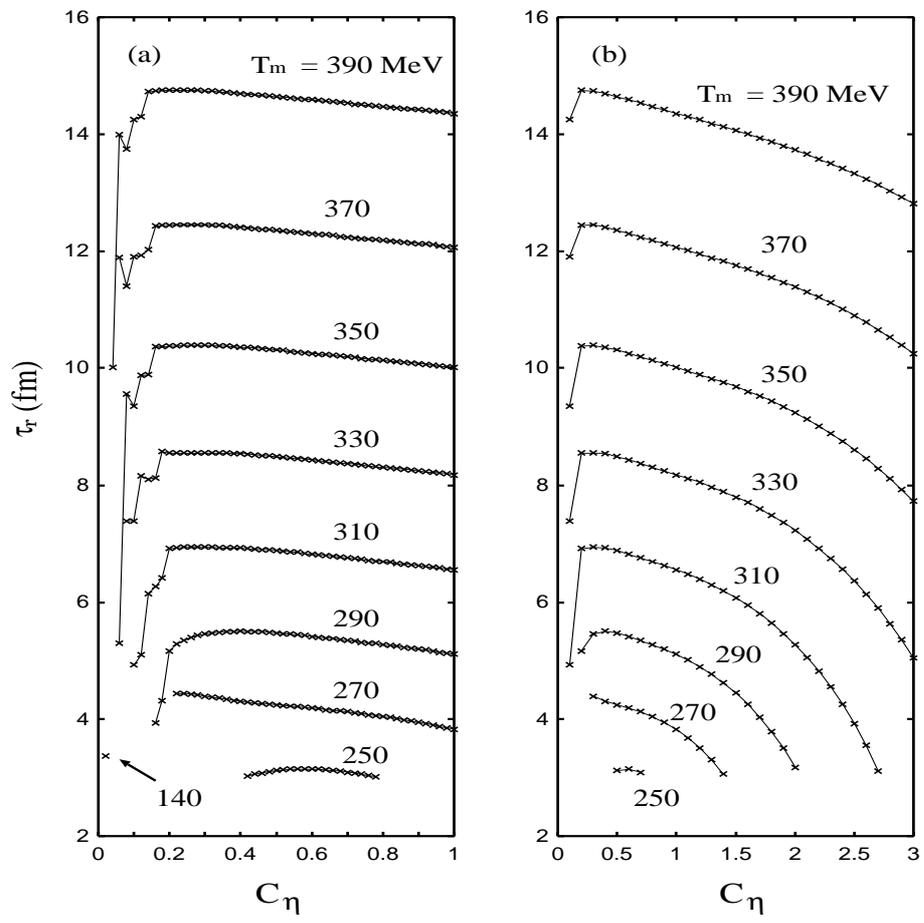
\begin{figure}
\caption{
The time interval ($\tau_{r}$) 
in which chiral symmetry restoration lasts longer than 3fm
for various $T_{m}$ with $T_{i} $ = 1 MeV and $\tau_{m}$ = 1 fm in  case A.
}
\label{fig:New_biro_time-tf1}
\end{figure}

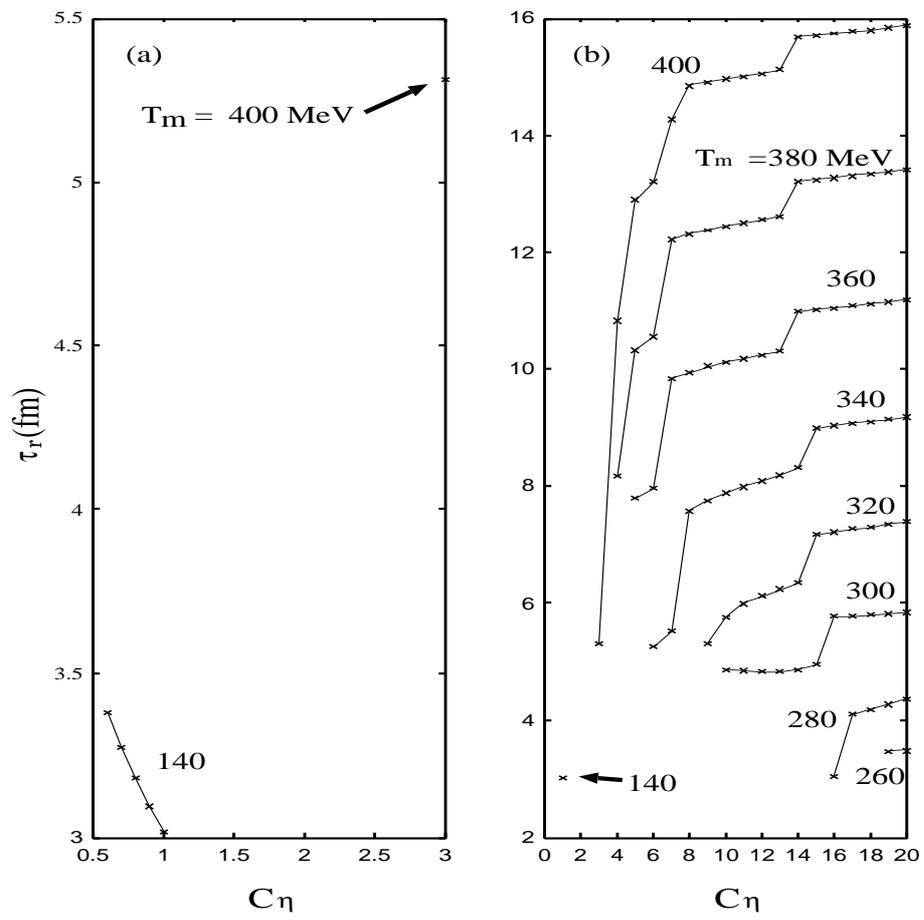
\begin{figure}
\caption{  
The time interval ($\tau_{r}$) in which chiral symmetry restoration lasts longer than 3fm
for various $T_{m}$ with $T_{i} $ = 1 MeV and $\tau_{m}$ = 1 fm in case B. 
it is displayed for (a) $0 \le C_{\eta} \le 1$ and (b) $0 \le C_{\eta} \le 3$.
}
\label{fig:m135_2}
\end{figure}

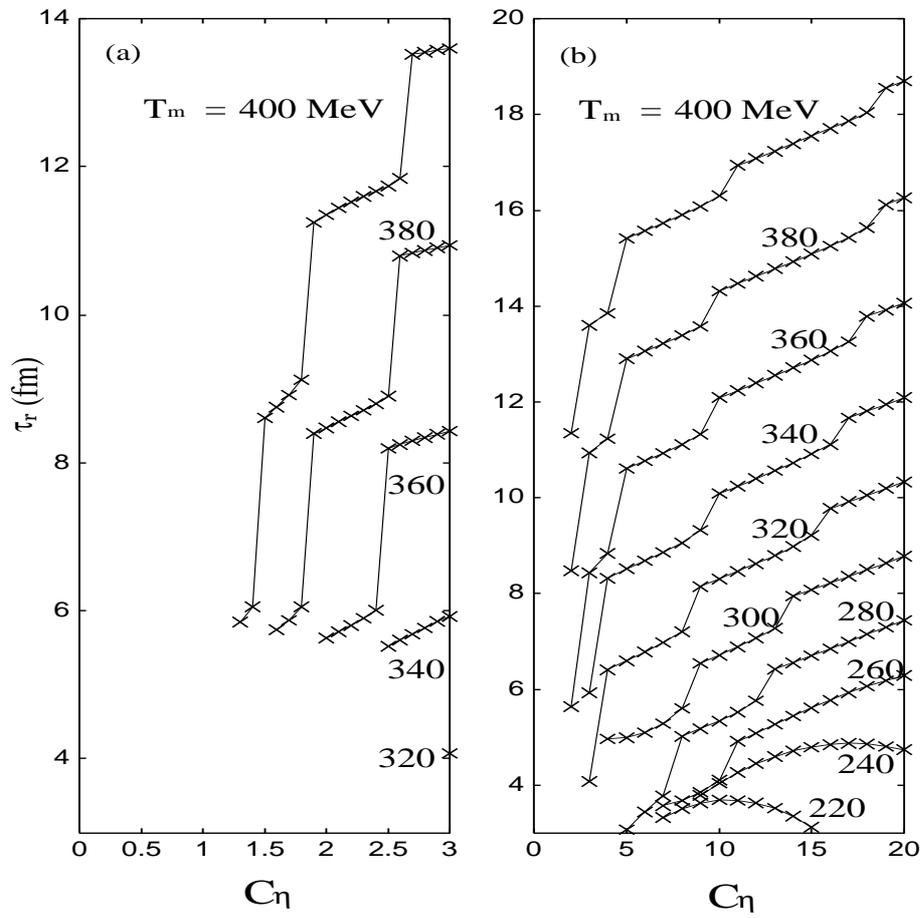
\begin{figure}
\caption{
The time interval ($\tau_{r}$) in which chiral symmetry restoration lasts longer than 3fm
for various $T_{m}$ with $T_{i} $ = 1 MeV and $\tau_{m}$ = 1 fm 
in case C.
it is displayed for (a) $0.5 \le C_{\eta} \le 3$ and (b) $0 \le C_{\eta} \le 20$.
}
\label{fig:effective2}
\end{figure}

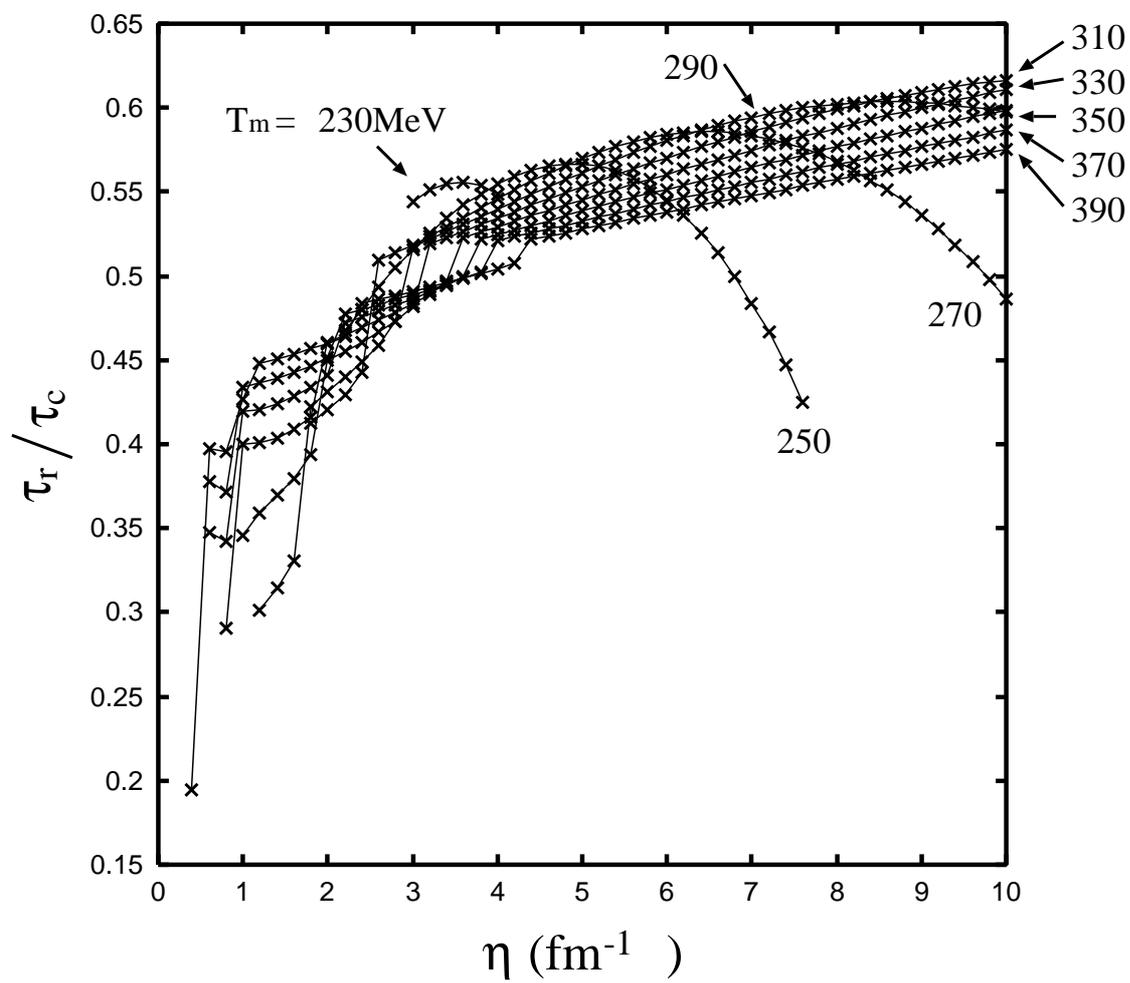
\begin{figure}
\caption{
The time interval ($\tau_{r}$) in which chiral symmetry restoration lasts longer than 3fm
for various $T_{m}$ with $T_{i} $ = 1 MeV and $\tau_{m}$ = 1 fm 
in case D.
it is displayed for (a) $0 \le C_{\eta} \le 3$ and (b) $0 \le C_{\eta} \le 20$.
}
\label{fig:VEV2}
\end{figure}

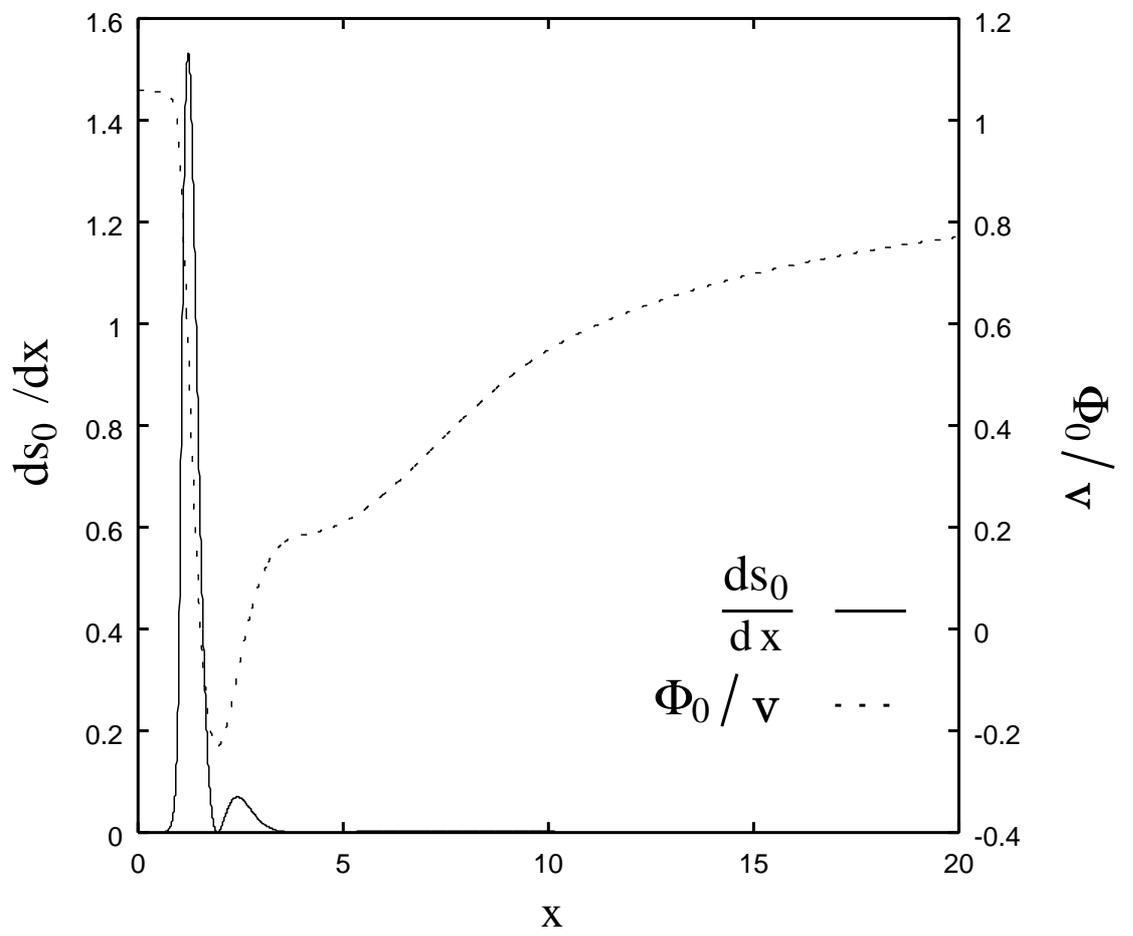
\begin{figure}
\caption{
The ratio $\tau_{r}/\tau_{c}$ for various $T_{m}$ in case A.
}
\label{fig:scale_const}
\end{figure}

\begin{figure}
\caption{
Entropy production per unit scaled time $(\tau/\tau_{m})$ without the freezeout 
with $T_{m}$ = 250 MeV , $T_{i}$ = 1 MeV, $\tau_{m}$ = 1fm and $\eta = 2 {\rm fm}^{-1}$
in case A.
}
\label{fig:ds0dx}
\end{figure}
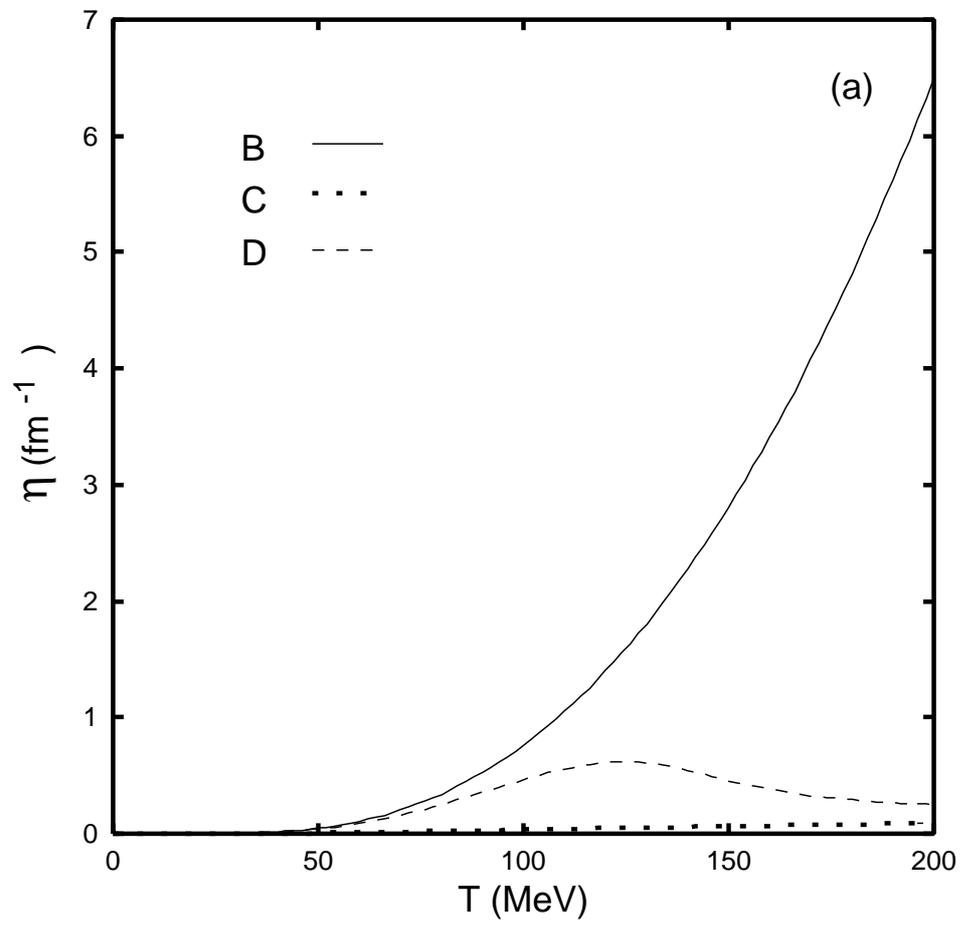
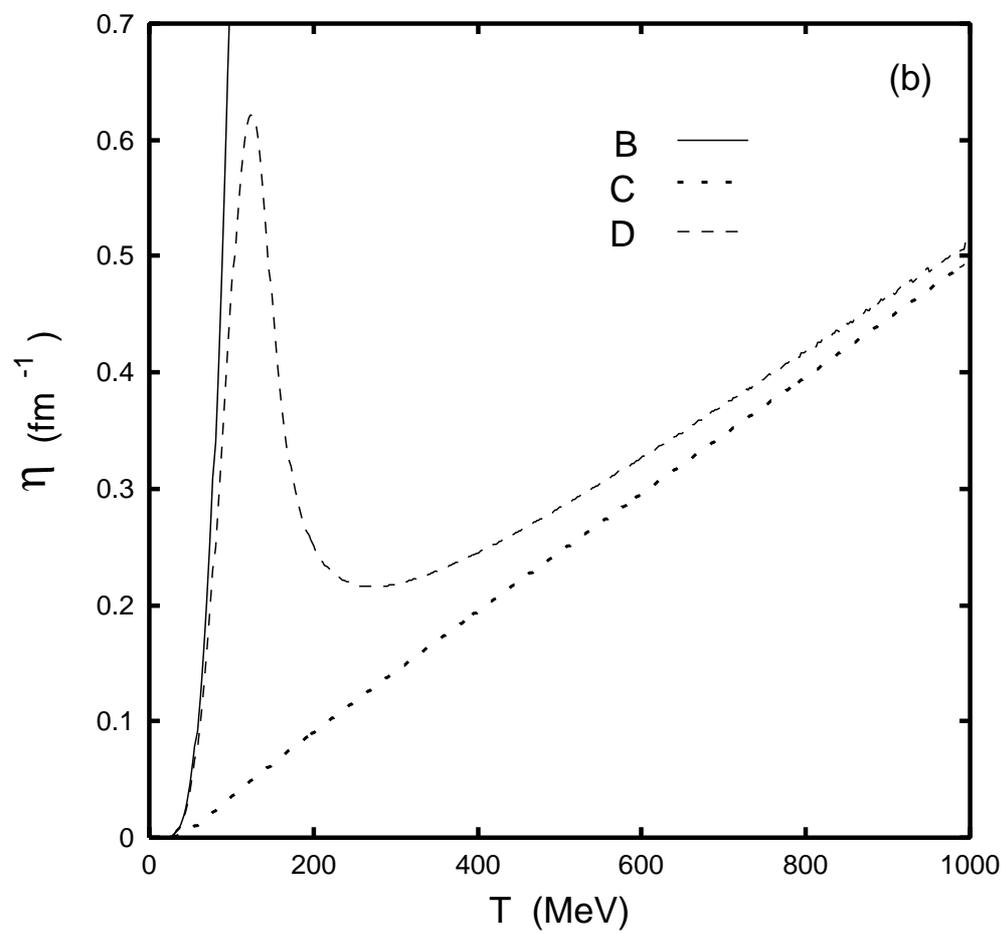
\end{document}